\documentclass[twocolumn]{aastex631}

\begin{document}

\title{Compact, AGN-hosting Dwarf Galaxies with ``Little Red Dot''-like SEDs in the Local Universe}

\author[0009-0009-6319-0888]{Lulu Bao}
\affiliation{University of Chinese Academy of Sciences, Beijing 100049, China}
\affiliation{State Key Laboratory of Radio Astronomy and Technology, National Astronomical Observatories, CAS, Beijing 100101, China}

\author[0000-0002-9390-9672]{Chao-Wei Tsai}
\affiliation{State Key Laboratory of Radio Astronomy and Technology, National Astronomical Observatories, CAS, Beijing 100101, China}
\affiliation{Institute for Frontiers in Astronomy and Astrophysics, Beijing Normal University, Beijing 102206, China} \affiliation{University of Chinese Academy of Sciences, Beijing 100049, China}
\email{cwtsai@nao.cas.cn}

\author[0000-0001-7808-3756]{Jingwen Wu}
\affiliation{University of Chinese Academy of Sciences, Beijing 100049, China}
\affiliation{State Key Laboratory of Radio Astronomy and Technology, National Astronomical Observatories, CAS, Beijing 100101, China}

\author[0009-0001-9170-3363]{Jialai Wang}
\affiliation{Department of Astronomy, University of Science and Technology of China, Hefei 230026, China}
\affiliation{School of Astronomy and Space Science, University of Science and Technology of China, Hefei 230026, China}



\begin{abstract}

Local active galactic nuclei (AGNs) in dwarf galaxies are often considered as analogs for the earliest supermassive black holes, although their connections require more comprehensive examinations. Motivated by finding the local analogs of ``Little Red Dots'' (LRDs), the compact, red galaxies discovered by \emph{JWST} and at $z > 5$ characterized by ``V-shaped'' SEDs, we compile a sample of local AGN-hosting dwarf galaxies (ADGs) with comparable luminosities to statistically evaluate this connection. By applying K-means clustering to SED shapes and morphological sizes, we classified four groups which trace a sequence in physical properties, including metallicity, star formation rate, and dust emission, mainly driven by their distinct UV–optical slopes. Within these groups, we find that about half of the ADGs exhibit ``V-shaped'' SEDs and relatively compact morphologies. However, a direct comparison reveals fundamental physical differences: local ``V-shaped'', compact ADGs appear significantly more evolved than high-$z$ LRDs, characterized by systematically larger effective radii and distinct ionization states. Our results suggest that local compact ADGs likely follow a different formation pathway from LRDs, highlighting the complexity of black hole-galaxy co-evolution across cosmic time.

\end{abstract}



\section{Introduction} \label{sec:intro}
Dwarf galaxies in the local Universe are often regarded as analogs of high-$z$ galaxies due to their low stellar masses ($M_{*} < 10^{9.5}\, M_{\odot}$), intense star formation, and low metallicities, particularly in some compact subsamples (e.g. \citealt{Cardamone2009,Amorin2010,yang2017}). 
Compared to massive galaxies, these low-mass galaxies with small super massive black holes (SMBHs) have a relatively poor merger and accretion history \citep{Reines2016}. As a result, the local AGN-hosting dwarf galaxies (ADGs) can provide special insights into the black hole (BH) seeding mechanism, the early growth of the BH and its coevolution with the host galaxy \citep{greene2020,Inayoshi2020,Reines2022}, which has become increasingly relevant with the increasing number of AGNs now being identified at high redshift (e.g. \citealt{Fan2023,greene2024uncover}).

It is challenging to identify the AGN activities in the dwarf galaxies \citep{greene2020,Reines2022}. Over the past decade, ADGs have been selected through multiple techniques, including the BPT (``Baldwin, Phillips \& Terlevich''; \citealt{bpt1981,bpt1987}) emission-line diagnostic diagrams \citep{Reines2013,Mezcua2020,Molina2021}, X-ray detection \citep{Lemons2015,Pardo2016,Mezcua2018,Birchall2020,Latimer2021,bykov2024srg}, and optical variability methods \citep{Baldassare2018,Baldassare2020,ward2022variability}.
These ADGs typically exhibit luminosities of $\sim 10^{12}\,L_{\odot}$ and black hole masses of $\sim 10^{6-7}\,M_{\odot}$ \citep{Reines2013,Reines2015}, located at low-mass end of the SMBH–host galaxy scaling relations.

The James Webb Space Telescope (\emph{JWST}) has revealed a large population of galaxies at redshifts $z>5$, revolutionizing our understanding of galaxy evolution and BH growth \citep{matthee2024little,kocevski2024rise}.
Among these, the so-called ``Little Red Dots'' (LRDs) are a population of compact, point-like galaxies with rest-frame UV–optical spectral energy distributions (SEDs) characterized by excess UV emission and a red optical continuum (referred to as the ``V-shaped'' SED feature;\,\citealt{matthee2024little,greene2024uncover,Labbe2023uncover,kocevski2024rise,williams2024galaxies}). LRDs are unexpectedly abundant, with number densities of $\sim 10^{-5}\, \mathrm{Mpc}^{-3}\,\mathrm{dex}^{-1}$, which is significantly higher than pre-\emph{JWST} predictions for quasar number density \citep{kokorev2024census,greene2024uncover}, posing challenges to current theories of galaxy evolution.

The physical nature of LRDs remains under debates. Their red rest-frame optical emission is often attributed to AGNs, consistent with the detection of broad Balmer emission lines in majority of sources \citep{Labbe2023uncover,matthee2024little}. The origin of the UV excess has been proposed to arise either from scattered AGN light or from the host galaxy \citep{Labbe2023uncover,greene2024uncover,leung2024exploring,ma2025}. 
Alternative models, such as AGNs embedded within dense gaseous envelopes, have also been invoked \citep{lizhengrong2025,kido2025,naidu2025}.
\textcolor{black}{Observationally, non-detections in stacked X-ray images (\citealt{yue2024, ananna2024ApJ, Andrea2025, maiolino2405jwst}) and the lack of significant variability \citep{zhang2025variability} suggest that the central engines are either deeply buried or intrinsically weak.}
In addition, LRDs exhibit weak rest-frame near-infrared (NIR) emission \citep{akins2024cosmos} and faint sub-millimeter fluxes (\citealt{akins2024cosmos}; \citealt{Xiao2025}; \citealt{Casey2025}), indicating a suppressed host-galaxy contribution. Comparisons with Hot Dust-Obscured Galaxies (Hot DOGs) further highlight the intrinsic differences between high-$z$ LRDs and dusty-obscured AGNs mostly at the cosmic noon \citep{bao2025}.
Furthermore, LRDs host overmassive SMBHs compared to the local BH–host galaxy relation \citep{greene2024uncover}. The rapid assembly of such high-$z$ SMBHs may point to the super-Eddington accretion, merger events, or the descendants of massive BH seeds \citep{greene2024uncover}.

LRDs are primarily identified at $z\sim5$–7, yet their number density declines steeply toward lower redshifts ($z<5$; \citealt{Ma2025LF,Ma2025lowz}). 
Galaxies exhibiting compact morphologies and V-shaped SEDs are exceedingly rare during the Cosmic Noon epoch ($z\sim2$–3). A few galaxies, such as the so-called ``Big Red Dots'' (\citealt{Loiacono2025}; \citealt{Stepney2024}) and the ``Rosetta Stone'' galaxy \citep{Rosettastone2024} with similar SED feature but larger effective radii, have been reported.
\textcolor{black}{In the local Universe, a small number of compact galaxies with ``V-shaped'' rest-frame UV-to-optical SEDs have been identified, including seven broad-line Green Pea galaxies \citep{linruqiu2025}, which have been suggested as potential local analogs to LRDs. Recent studies further indicate that these galaxies may exhibit weak, but non-zero, variability \citep{Lin2026GP}.
Other searches reveal that a small number of candidates with similar properties have been identified \citep{lin2025local,Ding2026}.} Only few nearby sources exhibit UV-to-optical SEDs, weak X-ray emission, and emission-line properties indicative of gaseous structures resembling those of high-$z$ LRDs \citep{lin2025local,ji2025local}.
The different number density across the cosmic time raise the question whether LRDs represent a evolutionary phase or only specific product at high redshifts.

AGNs in dwarf galaxies may preserve key physical conditions relevant to early black hole growth. In this work, we investigate the SED and morphological properties of a large compiled sample of local ADGs that exhibit comparable luminosities and black hole masses to high-$z$ LRDs, aiming to provide insights into their possible formation pathways.
The paper is organized as follows. In Section 2, we describe the construction of the LRD sample and the collection of ADGs. Section 3 presents the measurements and the classification method applied to the dwarf galaxy sample. In Section 4, we compare and discuss the statistical properties between compact ADGs and the high-$z$ LRDs. Finally, Section 5 summarizes our conclusions.
Throughout this paper, we adopt the AB magnitude system and assume a flat $\Lambda$CDM cosmology with $H_0 = 70~\mathrm{km~s^{-1}~Mpc^{-1}}$ and $\Omega_{M} = 0.3$ \citep{Hinshaw2013}.

\section{Galaxy sample} \label{sec:style}

\subsection{AGN-Hosting Dwarf Galaxies}
The sample of ADGs is compiled from 45 published studies up to the end of 2023 that report either AGN-host dwarf galaxies or massive black hole hosts at redshifts $z \lesssim 1$, following the compilation of Wang et al. (to be submitted). 
After applying uniform SED fitting with \texttt{CIGALE} \citep{cigale2019} to isolate the dwarf galaxy population, the final sample consists of 1,153 galaxies with stellar masses $M_* < 10^{9.5}\,M_{\odot}$. 
These galaxies were identified using a combination of selection techniques, including broad H$\alpha$ emission (12\%), optical emission-line diagnostics such as the BPT diagram (21\%), X-ray emission (14\%), optical variability (24\%), the detection of high-ionization emission lines such as [Fe\,\textsc{X}] and [He\,\textsc{ii}] (10\%), and other methods such as infrared color or radio excess (20\%). 
The details about this sample can be found in Wang et al. (to be submitted).

In addition, we incorporated dwarf galaxies hosting AGNs reported by \citet{Pucha2025}, spectroscopically identified in the DESI Early Data Release (EDR) and 20\% of Year 1 (DA0.2) data. We only included broad-line galaxies with stellar masses below $10^{9.5}\,M_{\odot}$ (estimated with \texttt{CIGALE}) and classified as AGN or composite in the BPT diagram. After cross-matching with \emph{GALEX} photometry, the DESI-identified dwarf galaxy subsample contains 53 objects.

\begin{figure*}[!htbp]
    \centering
    \includegraphics[width=0.9\linewidth]{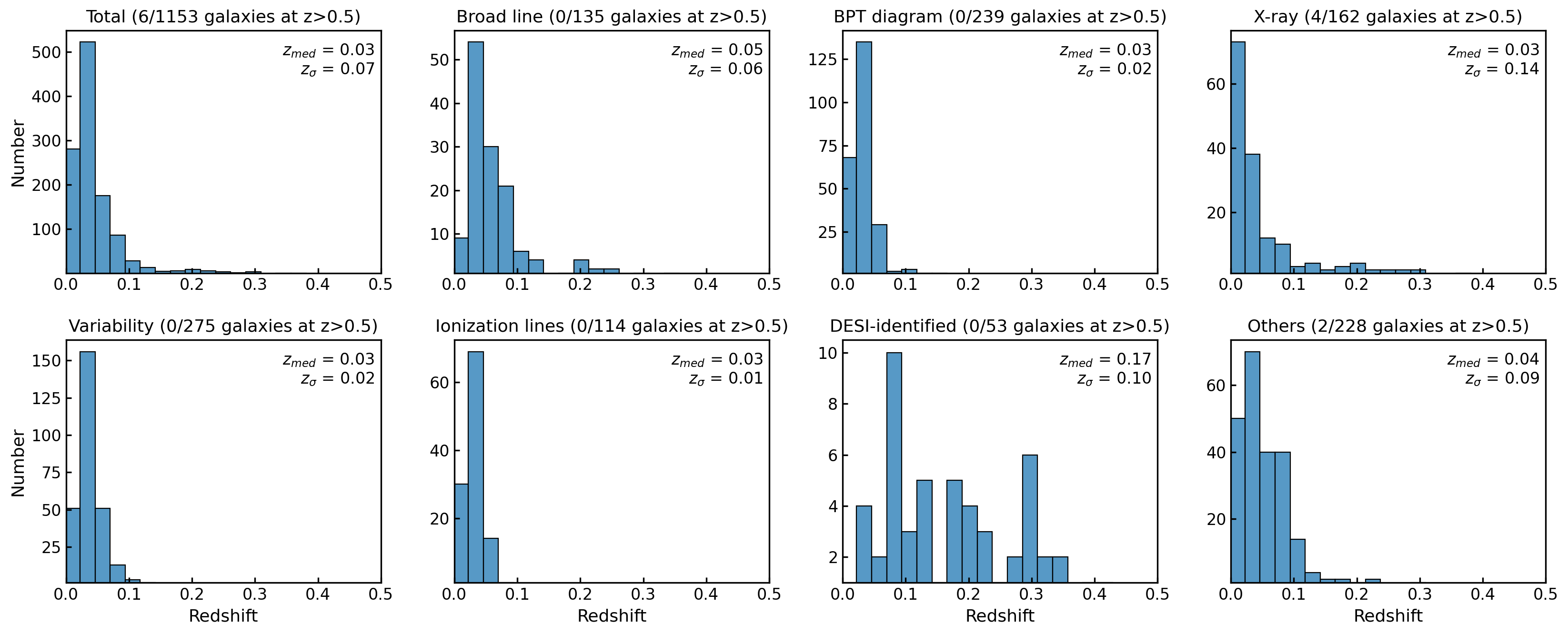}\\ 
    \caption{Redshift distributions of the total ADG sample and its subsamples (selected via broad H$\alpha$ emission, BPT diagnostics, X-ray emission, optical variability, high-ionization emission lines, DESI spectroscopy, and other methods such as infrared or radio emission). 
    The fraction of galaxies with $z>0.5$ in each sample (or subsample) is indicated at the top of each panel. 
    The median redshift and the $1\sigma$ scatter of each (sub)sample are also shown in the corresponding panels.
    }
    \label{redshift}
\end{figure*}

The redshift distribution of the total ADG sample and its subsamples are shown in Figure. \ref{redshift}. Given that the median redshift of the total sample is low ($\sim 0.03$) and few sources extend to higher redshifts ($> 0.5$), the ratio of galaxies with $z > 0.5$ to the total number of sources in each (sub)sample is marked at the top of each panel for comparison. Notably, the galaxies selected by DESI spectra show systematic higher redshift.

We obtained multi-wavelength photometric data of a total sample of 1,204 ADGs from sky surveys. Optical imaging data and photometry were obtained from the DESI Legacy Imaging Surveys Data Release 9 (DESI LS DR9; \citealt{DESI2019}) and the Sloan Digital Sky Survey Data Release 18 (SDSS DR18; \citealt{SDSSdr18-2023}). We adopted the SDSS cModel magnitudes to ensure reliable measurements of the total light for extended local sources. The Pan-STARRS photometry was not included, as the SDSS $u$-band data are essential for better constraining the UV slope for galaxies at $z>0.025$.
Infrared photometry was taken from the CatWISE2020 catalog \citep{catwise2021} (W1 and W2 at 3.4 and 4.6 $\mu$m) and the AllWISE catalog \citep{allwise2014} (W3 and W4 at 12 and 22 $\mu$m), both based on the Wide-field Infrared Survey Explorer (WISE) mission \citep{wise2010}.
Ultraviolet photometry in the near-UV (NUV) and far-UV (FUV) bands was cross-matched from the Galaxy Evolution Explorer (\emph{GALEX}; \citealt{galex2005}) survey.
Optical spectra were obtained from the SDSS spectroscopic database, which provides broader coverage for the ADG sample.

\subsection{LRDs collection}

We compiled photometric data of LRDs from \citet{akins2024cosmos} and \citet{kokorev2024census}. \citet{akins2024cosmos} reported 434 sources from the COSMOS-Web program \citep{Casey2023} and the PRIMER survey \citep{dunlop2021primer}, selecting objects based on a compactness metric in F444W, $0.5 < f_{\rm F444W}(d=0''.2)/f_{\rm F444W}(d=0''.5) < 0.7$, and a red F277W–F444W color, $m_{\rm F277W} - m_{\rm F444W} > 1.5$. This single-color criterion differs from other LRD selections using red rest-frame optical and blue UV colors \citep[e.g.,][]{Labbe2023uncover, kocevski2024rise} and is more similar to the selection of extremely red objects (EROs) \citep{akins2024cosmos}.  
From this sample, we selected 76 “V-shaped” sources satisfying the ``\emph{red2}'' criteria of (\citealt{Labbe2023uncover}; $m_{\rm F150W} - m_{\rm F200W} < 0.8$, $m_{\rm F277W} - m_{\rm F356W} > 0.7$, and $m_{\rm F277W} - m_{\rm F444W} > 1.0$). \citet{kokorev2024census} reported 260 LRDs using similar ``V-shaped'' criteria from CEERS \citep{Finkelstein2022ceers}, PRIMER \citep{dunlop2021primer}, and other programs covering GOODS-S \citep{Giavalisco2004}.  

In this work, we focus on these 326 LRDs, all showing a ``V-shaped'' feature from rest-frame UV to optical. Both studies provide effective radii $R_e\lesssim100\,$pc measured in the NIRCam/F444W band. We also adopt the photometric data and stacked SED from \citet{akins2024cosmos} as a template to quantify SED similarity between LRDs and ADGs.

\section{K-means clustering analysis and comparisons}

\subsection{Line emission, photometric, and morphological properties}

The emission-line fitting package \texttt{PyQSOFit} \citep{Guo2018} was used to model the continuum and emission lines of the SDSS spectra for the ADG sample. \texttt{PyQSOFit} decomposes the host-galaxy stellar continuum and the AGN power-law emission, upon which individual emission lines are fitted. The fitted lines include [O\,\textsc{iii}]~$\lambda5007$, H$\alpha$, H$\beta$, [N\,\textsc{ii}]~$\lambda\lambda6548,6583$, and [S\,\textsc{ii}]~$\lambda\lambda6716,6731$. A total of 970 galaxies with available SDSS spectra were analyzed. Broad Balmer emission lines (FWHM $>$ 1200~km\,s$^{-1}$) are detected in only $\sim$8\% of the galaxies in our sample. 

The UV and optical continuum slopes of both the LRD and dwarf galaxy samples were measured following the method of \citet{kocevski2024rise}, applied to both the LRD and dwarf galaxy samples.
For each galaxy, photometric bands were divided into two groups according to whether their rest-frame effective wavelengths, determined by the source redshift, lie blueward or redward of the Balmer break.
For JWST-detected sources, this selection was drawn from HST/WFC3 F814W and JWST/NIRCam filters (F090W, F115W, F150W, F200W, F277W, F356W, F410W, and F444W).
\textcolor{black}{For the local sample, the same rest-frame wavelength criterion was used to divide the SDSS $u,g,r,i,z$ and GALEX NUV/FUV bands into blueward and redward groups.}
The slopes were then computed as $\beta = 0.4(m_1-m_2)/\mathrm{log}(\lambda_2/\lambda_1)-2$, where $m_1$ and $m_2$ are AB magnitudes at effective wavelengths $\lambda_1$ and $\lambda_2$.
Uncertainties in the slopes were estimated by propagating the photometric uncertainties through Monte Carlo simulations.
For each band, we performed 1,000 realizations by resampling the observed fluxes based on their photometric errors, repeated the slope measurements, and adopted the standard deviation of the resulting distribution as the final uncertainty. 

\begin{figure}[!htbp]
    \centering
    \includegraphics[width=0.9\linewidth]{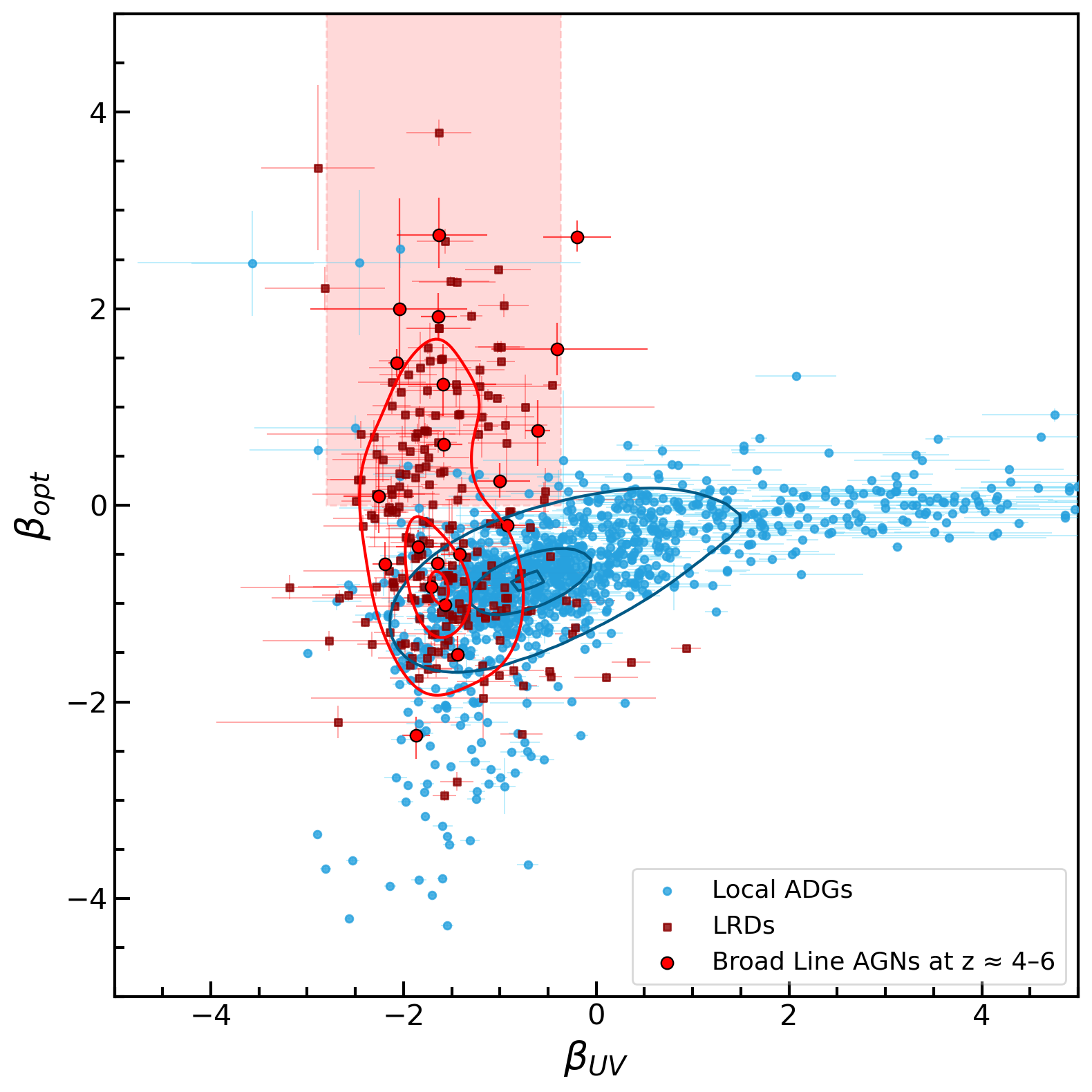}\\ 
    \caption{Distribution of rest-frame UV and optical continuum slopes for LRDs and AGN-hosting dwarf galaxies (ADGs). Red squares denote LRDs from \citet{akins2024cosmos} and \citet{kokorev2024census}, while blue dots represent ADGs. Broad H$\alpha$ line emitting AGNs from \citet{matthee2024little} are shown as red points. The solid red and dark-blue curves indicate the 1$\sigma$ and 2$\sigma$ density contours for LRDs and ADGs, respectively. The red shaded region marks the ``V-shaped'' selection window for LRDs defined by \citet{kocevski2024rise}.
    }
    \label{slope}
\end{figure}

\textcolor{black}{To evaluate the impact of emission-line contamination on the photometry used to derive continuum slopes, we apply a line-subtraction test to the subset of local ADGs with available SDSS spectra.
Considering the contributions from [O\,\textsc{iii}] $\lambda5007$, H$\alpha$, and H$\beta$ measured by \texttt{PyQSOFit}, the $g$, $r$, and $i$ bands are affected. The induced change in optical slope results in a shift of $0.02^{+0.13}_{-0.02}$, which is comparable to the median uncertainty of $\beta{\mathrm{opt}}$ ($\sim 0.02$). Ly$\alpha$ may contaminate FUV photometry only for galaxies at $z \gtrsim 0.1$, comprising $\sim$5\% of the ADG sample. Given the minor impact, and for consistency with LRD photometry and UV photometry of ADGs, which generally lack spectra for systematic line subtraction, we adopt the uncorrected slopes in the main analysis.}
The UV and optical slopes of LRDs and dwarf galaxies are shown in Figure.~\ref{slope}. The red shaded region indicates the color-selection criterion defined by \citet{kocevski2024rise}, namely $\beta_{\rm opt} > 0$ and $-2.8 < \beta_{\rm UV} < -0.4$. Red dots correspond to LRDs, and blue dots represent ADGs. 

Compactness in morphology is another defining property of LRDs, with typical effective radii of $\sim$100 pc \citep{akins2024cosmos, kokorev2024census, Labbe2023uncover}. To compare the compactness of ADGs and LRDs, we measure the morphological properties of the ADG sample by modeling the surface brightness profiles using \texttt{GALFIT} \citep{Peng2002, Peng2010} on DESI LS DR9 imaging data, which offer seeing-limited ground-based but relatively deep optical images. 
\textcolor{black}{As previous morphological analyses of ADG host galaxies, conducted on subsets of our ADG sample, have shown that most sources exhibit bulge or pseudobulge structures in HST imaging \citep{Jiang2011,Schutte2019,Kimbrell2021}, we adopt a three-component model as our default, consisting of two Sérsic components (an extended host component and an inner high-concentration component) and one PSF to represent unresolved AGN emission. The inner component is not strictly interpreted as a classical bulge, but is instead treated as a compact structure potentially associated with the central AGN for the purpose of size characterization. For cases where the three-component fit yields unstable solutions, we adopt a fallback model with either two components (one Sérsic plus one PSF; 29 ADGs) or a single PSF (5 ADGs). Objects that are strongly affected by multi-source blending are excluded from the morphological analysis. With this fitting procedure, except for sources lacking usable DESI imaging coverage, we obtain reliable morphological measurements for 533 ADGs.
As a complementary check, we also compute a non-parametric compactness metric defined as the ratio of the half-light radius to the PSF FWHM ($R_{1/2}/\mathrm{FWHM}_{\rm PSF}$), and find a qualitatively similar size contrast between ADG groups, although this approach does not provide structural decomposition.}
Because the band-to-band scatter in effective radius (across $g$, $r$, and $z$) is much smaller than the inter-object variation, \textcolor{black}{we adopt the effective radius from the best-fitting band as the characteristic size of each ADG, using the higher-$n$ inner Sérsic component in three-component fits, the single Sérsic component in two-component fits, or the PSF size for unresolved sources.} The resulting sample has a median effective radius of $R_e \sim 2100 \pm 1200$ pc and median Sérsic index $n \sim 1.9 \pm 1.0$, substantially larger than the typical LRD size scale ($R_e \sim 100$ pc; \citealt{akins2024cosmos,kokorev2024census}).

\subsection{K-means clustering}

We applied an \textcolor{black}{unsupervised} K-means clustering analysis to the dwarf galaxy sample to identify sub-populations with similar physical and photometric properties. The analysis was implemented using the \texttt{Scikit-learn} package \citep{scikit2011}. The clustering was based on four parameters: the UV and optical continuum slopes ($\beta_{\mathrm{UV}}, \beta_{\mathrm{opt}}$), the effective radius ($R_e$), and the $\mathrm{FUV}-z$ color. \textcolor{black}{Specifically, the effective radius characterizes the galaxy morphology, while the two slopes describe the SED shapes in the UV and optical regimes, respectively. The $\mathrm{FUV}-z$ color, spanning a much broader wavelength range, traces the overall UV-to-optical contrast of the SED.}
To determine the optimal number of groups, we adopted the Elbow method and verified the result with the silhouette coefficient. Both criteria consistently indicated $k=4$ as the most appropriate choice. 

\textcolor{black}{Consequently, the sample was divided into four groups, comprising approximately 47\%, 29\%, 14\%, and 10\% of the galaxies, respectively. Based on their distinct characteristics, these groups can be described as: (1) a ``Compact\,\&V-shape'' group and (2) a ``Diffuse\,\&V-shape'' group, both exhibiting UV–optical SED profiles broadly consistent with those of LRDs, yet differing in their effective radii; (3) a ``Blue'' group and (4) a ``Red'' group, representing systems with monotonically decreasing and increasing UV-to-optical luminosity, respectively.}

\begin{figure*}[!htbp]
    \centering
    \includegraphics[width=0.8\linewidth]{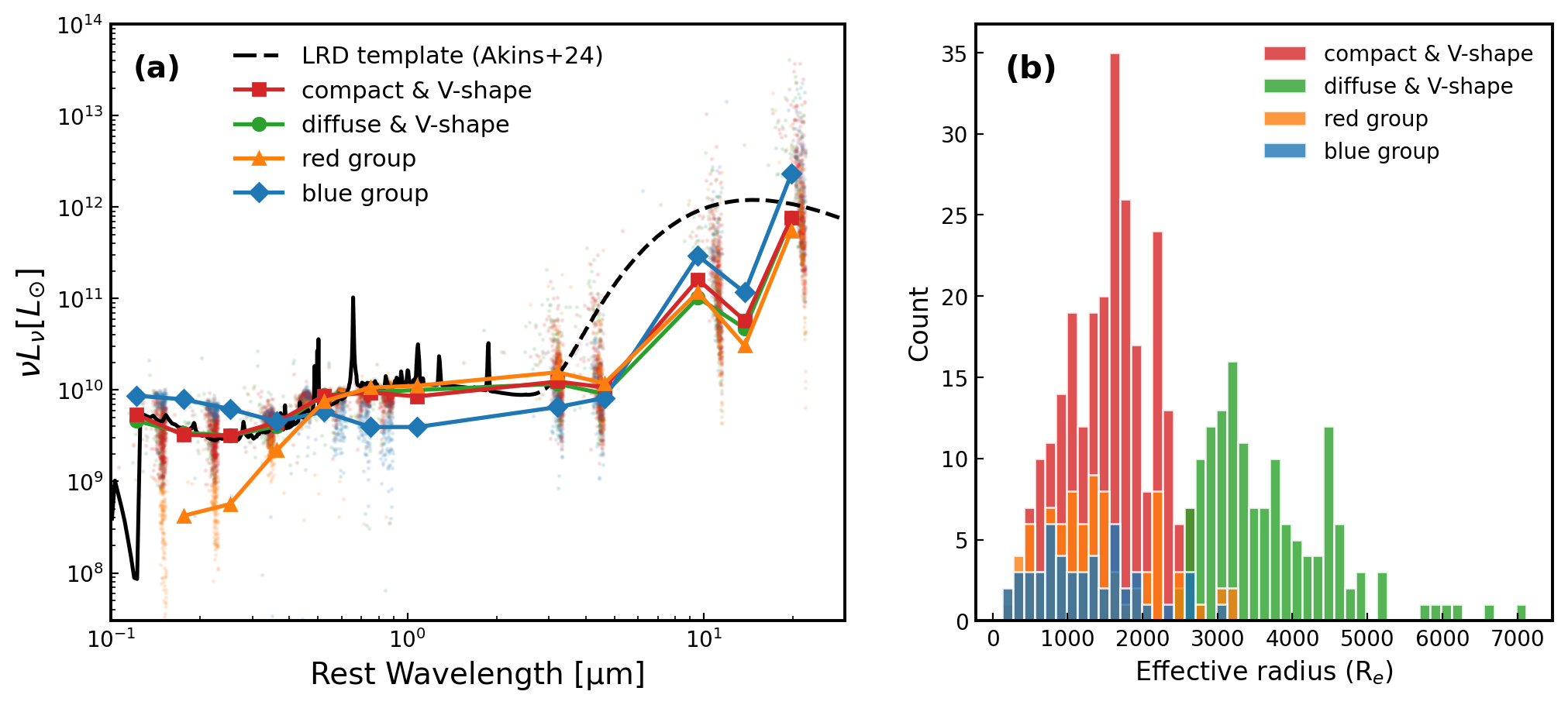}\\ 
    \caption{Stacked rest-frame SEDs (panel a) and effective radius ($R_e$) distributions (panel b) of the four ADG groups classified via the K-means clustering algorithm. In panel (a), the solid colored curves represent the floating-median SEDs for each group. Individual galaxy photometries are shown as semi-transparent scattered points, all of which are normalized to the luminosity of the LRD template from \citet{akins2024cosmos} (black curve). Redward of the observed-frame 18\,$\mu$m (corresponding to $\lambda_{\rm rest} \gtrsim 2.5\,\mu$m), the template is shown as a dashed line to indicate that it is constrained only by upper limits. In panel (b), the histograms illustrate the distribution of $R_e$ for each group, with color-coding consistent with panel (a).
    }
    \label{sed-group}
\end{figure*}

\textcolor{black}{
The stacked SEDs of the four groups are presented in the left panel of Figure~\ref{sed-group}. Each curve represents the floating–median SED constructed by scaling the SED of each galaxy to the luminosity of the LRD template before stacking. For comparison, the LRD template from \citet{akins2024cosmos} is shown as the black solid curve. 
The radius distributions of all groups are shown in the right panel of Figure~\ref{sed-group}.
While both the ``Compact\,\&V-shape'' and ``Diffuse\,\&V-shape'' groups show LRD-like SEDs, galaxies in the latter exhibit systematically larger effective radii.
We emphasize that the label ``Compact\,\&V-shape'' refers to galaxies that are relatively more compact ($R_e = 1.6^{+0.6}_{-0.7}$ kpc) compared with the ``Diffuse\,\&V-shape'' group ($R_e = 3.4^{+1.0}_{-0.6}$ kpc), with no absolute compactness threshold was imposed in the K-means clustering. In addition, the median effective radii of the ``Blue'' and ``Red'' group are $1.2^{+0.7}_{-0.6}$ kpc and $1.3^{+0.9}_{-0.6}$ kpc, respectively.
}

\begin{table*}[htbp]
\centering
\caption{Squared normalized differences between each ADG group and LRDs, including the normalized SED residuals relative to the LRD SED, UV and optical slopes, and effective radius. The median redshifts and the corresponding 16th--84th percentile ranges for each group are also shown.}
\label{tab:dwarf_lrd_diff}
\begin{tabular}{lccccc} 
\hline
Group & $z_{\rm med}$ & $\Delta \mathrm{SED_{residual}}^{2}$ & $\Delta \beta_{\mathrm{UV}}^{2}$ & $\Delta \beta_{\mathrm{opt}}^{2}$ & $\Delta \mathrm{R^{2}_e}$ \\
\hline
``Compact\,\&V-shape''    & $0.032^{+0.040}_{-0.017}$ & 0.8 & 0.6 & 1.2  & 2.7 \\
``Diffuse\,\&V-shape''    & $0.042^{+0.079}_{-0.015}$ & 1.1 & 0.3 & 0.8  & 10.8 \\
Blue group          & $0.038^{+0.041}_{-0.024}$ & 4.3 & 0.0 & 10.1 & 1.5 \\
Red group           & $0.027^{+0.018}_{-0.015}$ & 8.5 & 7.3 & 0.0  & 1.8 \\
\hline
\end{tabular}
\end{table*}

In addition to color–color space and effective radius, we quantify the similarity of overall SED shapes between LRDs and dwarf galaxies. To compare the overall characteristics of the two populations, we focused on four parameters: the characteristic ``V-shaped'' SED (the UV slope $\beta_{\mathrm{UV}}$ and the optical slope $\beta_{\mathrm{opt}}$), compact morphology (effective radius $R_e$), and the scaled SED residual. For each dwarf galaxy, the UV–optical photometry in seven bands was scaled to the luminosity of the LRD template from \citet{akins2024cosmos} using a least-squares fit. The scaled SED residual was then calculated as the root-mean-square error (RMSE) of the residuals across these seven bands. The same procedure was applied to the LRD photometry for consistency.
All four parameters were normalized, and Euclidean distances were computed between each ADG group and the normalized LRD property vector. We find that the ``Compact\,\&V-shape'' group with the smallest distance to the LRD properties corresponds to the largest group identified in our clustering analysis, comprising nearly half of the total sample. Table~\ref{tab:dwarf_lrd_diff} summarizes the squared normalized differences between the dwarf galaxies and LRDs, while a detailed comparison of the physical properties is presented in Section 3.3.

\subsection{Statistical properties of local ADG subsamples}
\begin{figure}[!htbp]
    \centering
    \includegraphics[width=0.9\linewidth]{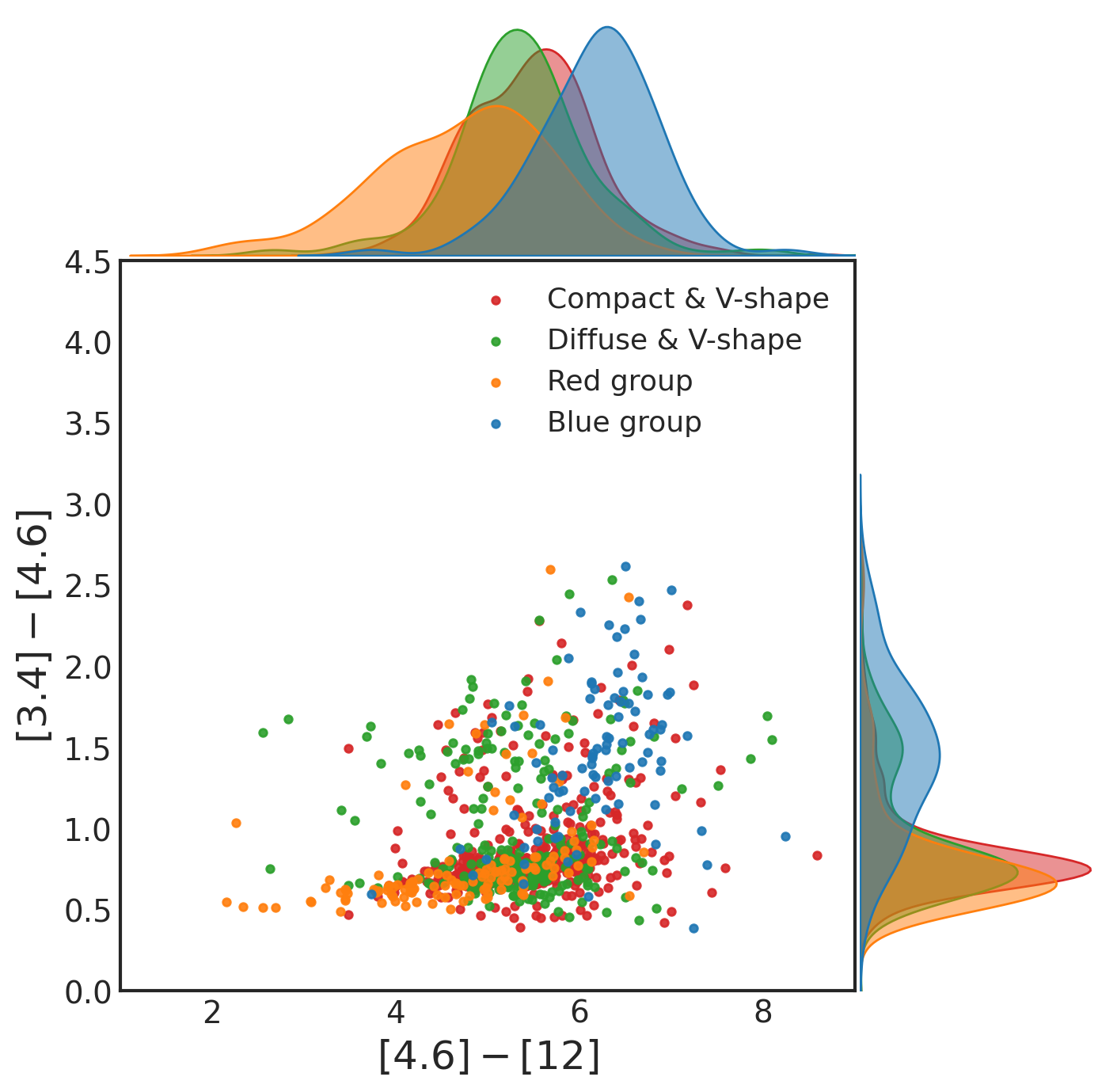}\\ 
    \caption{WISE color distribution of ADGs. Scatter points in different colors represent the groups classified by K-means. The marginal plots on the right and top show the number density distributions of W1$-$W2 and W2$-$W3 colors, respectively.}
    \label{wise-color}
\end{figure}

LRDs are rarely detected in the \emph{JWST}/MIRI bands \citep{akins2024cosmos,leung2024exploring}. The constraints on their dust properties thus remain limited. In contrast, the infrared emission of local ADG samples can be more readily investigated. We first investigate the dust emission using \emph{WISE} photometry. Distinct infrared properties are observed among the groups. As shown in Figure~\ref{wise-color}, the ``Blue''  group exhibits significantly redder W1$-$W2 and W2$-$W3 colors than the other groups. In contrast, the other three groups show comparatively bluer infrared colors.

\begin{figure*}[!htbp]
    \centering
    \includegraphics[width=0.7\linewidth]{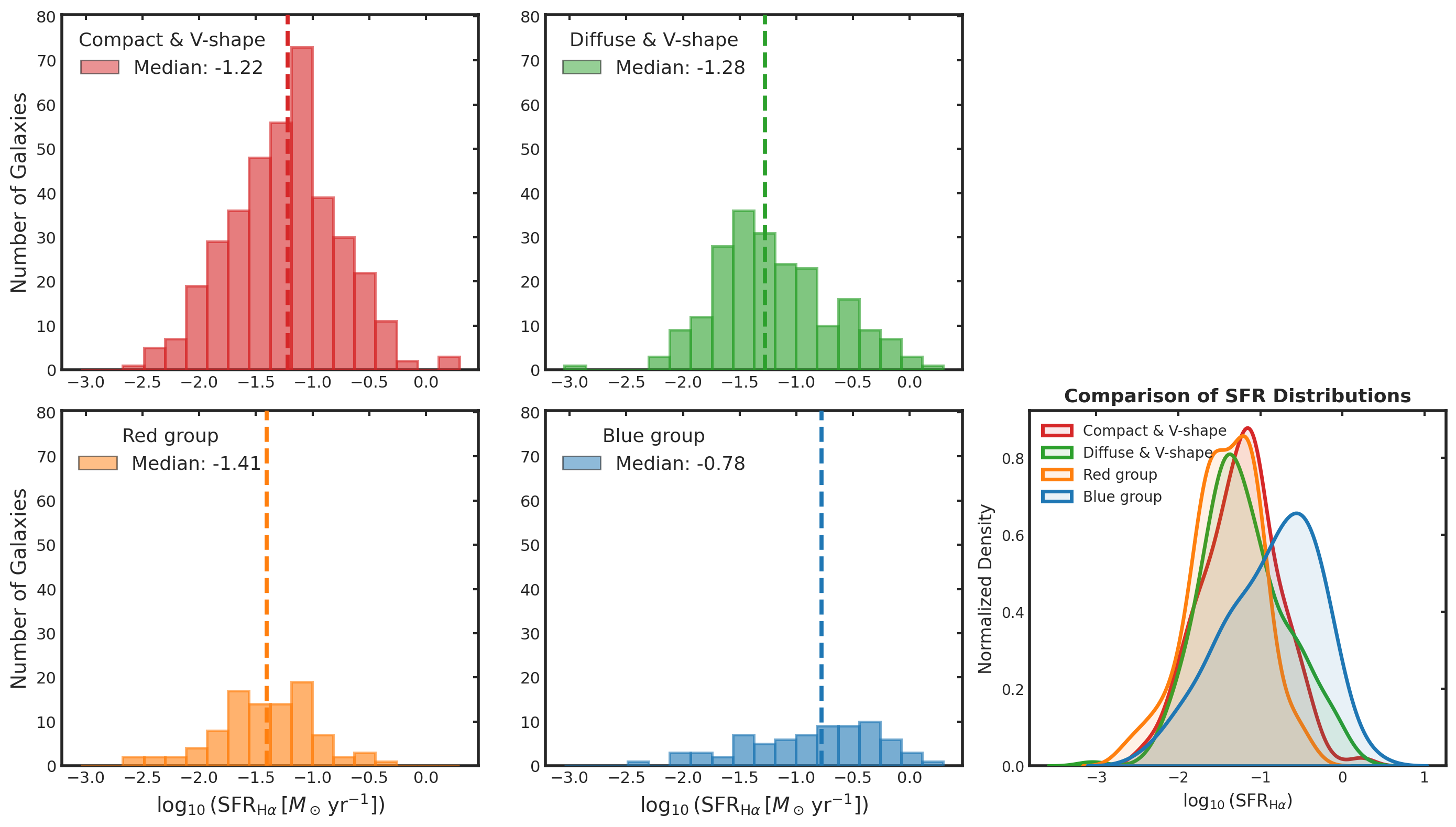}\\ 
    \caption{SFR distributions of the four ADG groups, derived from extinction-corrected narrow H$\alpha$ emission lines. In the four left panels, the vertical dashed lines indicate the median $\log_{10}(\mathrm{SFR}_{\mathrm{H}\alpha})$ for each group. The right panel displays the overlaid normalized density distributions. K-S tests confirm that the ``Blue'' group's SFR distribution is statistically distinct from all other groups ($p < 10^{-5}$).}
    \label{sfr}
\end{figure*}

We further investigate the star formation rates (SFRs) among the four groups of ADGs. Although both FUV luminosity and narrow H${\alpha}$ emission are widely employed as star formation tracers, the FUV-based estimates are subject to degeneracies associated with the UV slope that underlies our group classification. 
Using the standard calibrations described in \citet{Kennicutt1998}, we find that the FUV-derived SFRs are systematically higher than the H$\alpha$-derived values by $\sim 0.55$~dex. This discrepancy is consistent with the known breakdown of standard scaling relations in the dwarf galaxies \citep{Lee2009}. While both tracers may be partially contaminated by AGN activity, we adopt the extinction-corrected H$\alpha$ luminosity \citep[Eq.~10 of][]{Lee2009} as a more reliable tracer of recent star formation activity ($<10$~Myr) in these low-mass systems.
As shown in Figure~\ref{sfr}, the four groups exhibit clearly distinct SFR distributions. The ``Blue'' group has the highest median SFR of $\sim 0.17\, M_{\odot}\,\mathrm{yr^{-1}}$, while the ``Red'' group shows the lowest value of $\sim 0.04\, M_{\odot}\,\mathrm{yr^{-1}}$. The two groups characterized by V-shaped SEDs display intermediate star formation activity and show no statistically significant difference from each other ($p \approx 0.3$), as confirmed by Kolmogorov--Smirnov (K--S) tests.
Specifically, the ``Compact\,\&\,V-shape'' group has a median SFR of $\sim 0.06\,M_{\odot}\,\mathrm{yr^{-1}}$, while the ``Diffuse\,\&\,V-shape'' group shows a similar but slightly lower value of $\sim 0.05\,M_{\odot}\,\mathrm{yr^{-1}}$.

\begin{figure*}[!htbp]
    \centering
    \includegraphics[width=0.7\linewidth]{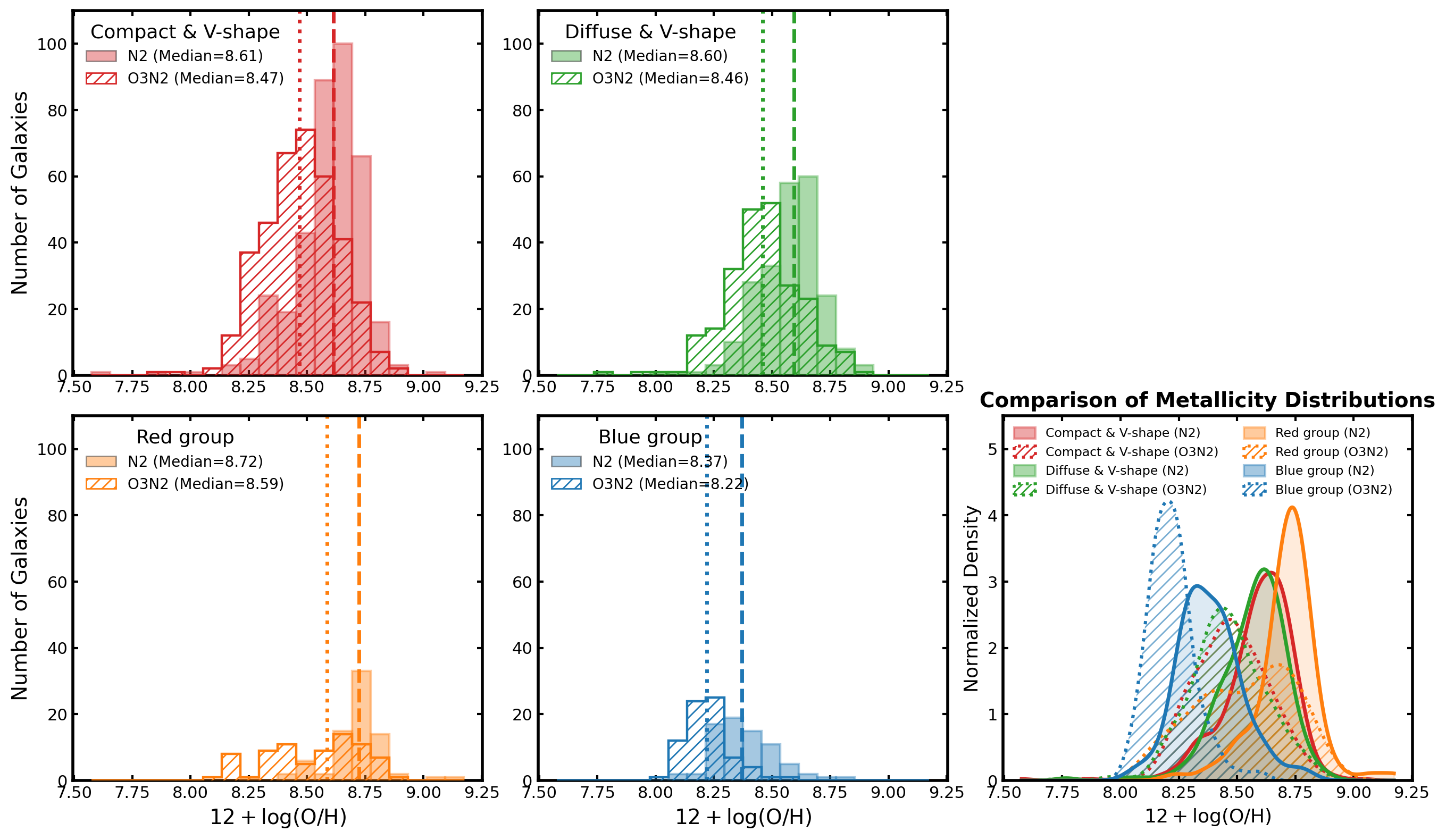}\\ 
    \caption{Gas-phase metallicity distributions of the four groups derived using the N2 (filled bars) and O3N2 (hatched bars) empirical calibrators. In the left four panels, the median metallicities obtained with each method are indicated by dashed (N2) and dotted (O3N2) vertical lines. The right panel presents the overlaid normalized density distributions measured with both indicators.}
    \label{metal}
\end{figure*}

We estimated the gas-phase metallicities of the dwarf galaxies using the empirical calibration of \citet{Pettini&Pagel}. The color excess $E(B-V)$ was derived from the H$\alpha$/H$\beta$ Balmer decrement using the \citet{Calzetti2000} reddening curve. Metallicity was then inferred from the [O\,\textsc{iii}] and [N\,\textsc{ii}] line diagnostics \citep{Pettini&Pagel}. We note that this metallicity estimate does not account for potential AGN contamination, which may introduce systematic uncertainties. Photoionization models show that the harder radiation field of an AGN produces higher ionization parameters than in typical H II regions and enhances high-ionization lines such as [O\,\textsc{iii}] $\lambda5007$, potentially biasing metallicity estimates based on [O \textsc{iii}] diagnostics \citep{Groves2006}. An inverse trend relative to the SFR is observed: galaxies with bluer colors tend to be more metal-poor. As shown in Figure~\ref{metal}, the four K-means groups exhibit clearly distinct metallicity distributions based on the N2 (which represents \,$\log([\mathrm{N\,II}]/\mathrm{H}\alpha)$\, index) and O3N2 (which represents \,$\log[([\mathrm{O\,III}]/\mathrm{H}\beta)/([\mathrm{N\,II}]/\mathrm{H}\alpha)]$\, index) calibrators. The ``Blue''  group is the most metal-poor; the two V-shape groups display intermediate metallicities with no statistically significant difference from each other ($p \approx 0.4$) by K-S tests; and the ``Red'' group is the most metal-enriched.

Based on the analysis presented in this subsection, the ``Blue'' group exhibits the bluest SED slopes, the most intense star formation, and the lowest metallicities and environmental densities, whereas the ``Red'' group shows the opposite trend. 
Despite their contrasting stellar populations, both groups remain relatively compact, with $R_e \lesssim 3$ kpc, indicating that the SED shape strongly influences many of their physical properties. 
In comparison, the two ``V-shaped'' groups occupy an intermediate regime between the Blue and Red groups, with their physical parameters showing transitional distributions.

\subsection{Environment properties}
\begin{figure*}[!htbp]
    \centering
    \includegraphics[width=0.7\linewidth]{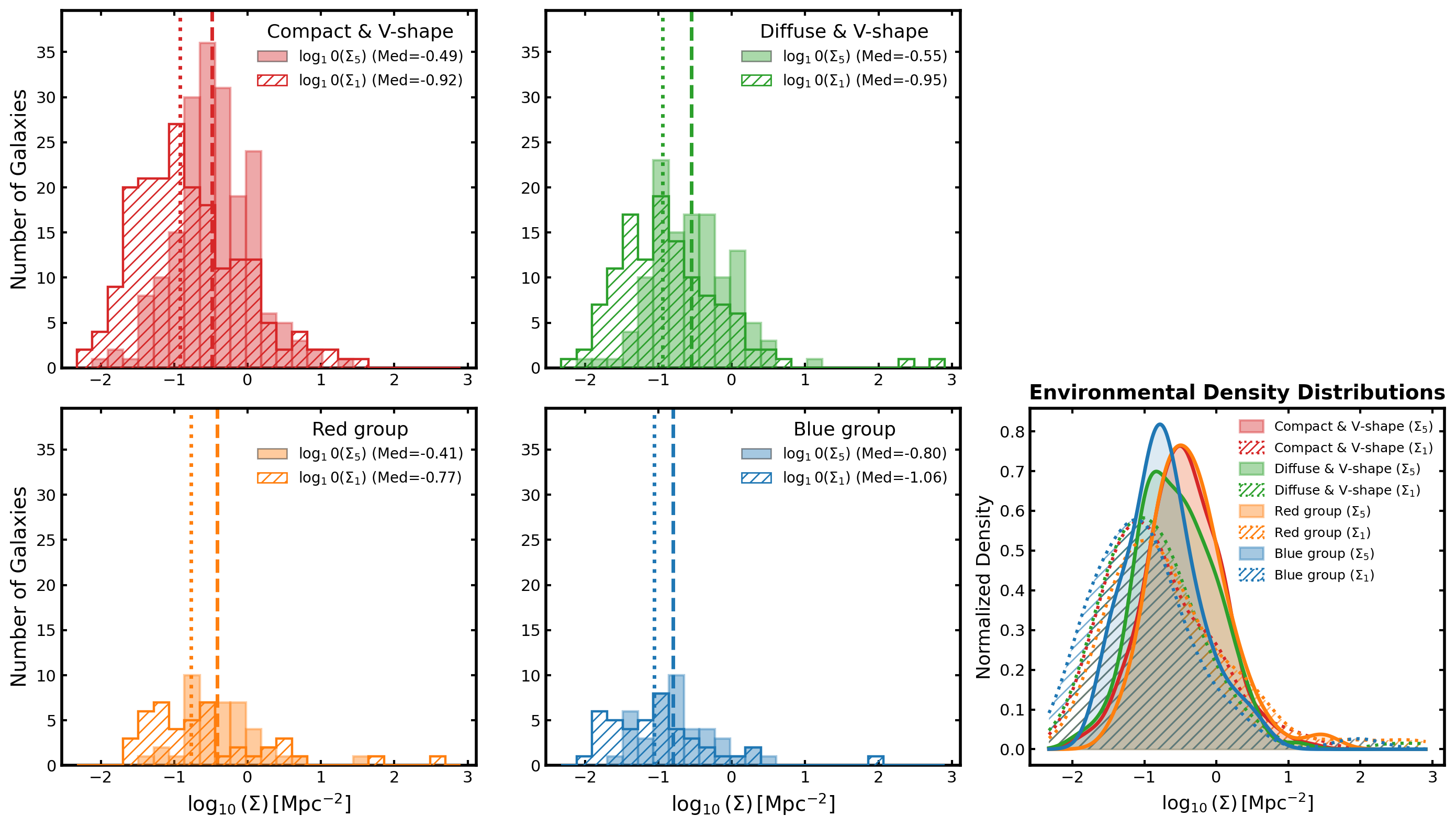}\\ 
    \caption{Distribution of the local galaxy surface density around each ADG in the four groups, estimated using the $k$NN method. In the left four panels, filled bars show the density traced by the fifth-nearest neighbors from spectroscopically identified DESI and SDSS galaxies with $M_r > -20.5$ and relative velocities within $5000\,\mathrm{km\,s^{-1}}$ ($\Sigma_5$), while hatched bars represent the density traced by the nearest massive galaxy with $M_r > -22.5$ under the same velocity cut ($\Sigma_1$). The median densities are indicated by dashed ($\Sigma_5$) and dotted ($\Sigma_1$) vertical lines. The right panel presents the overlaid normalized density distributions derived from the two $k$NN estimators.}
    \label{env}
\end{figure*}

To investigate whether the observed variations in SED shape, effective radii, and other properties are environmentally driven, we analyze the local environments of the four groups. We employ a $k$-nearest neighbor ($k$NN) approach, using galaxies with spectroscopic redshifts from SDSS DR17 \citep{SDSS17} and DESI DR1 \citep{DESIdr1} as tracers.
First, we calculate the surface density based on the 5th-nearest neighbor ($\Sigma_5$), selecting companions with absolute magnitudes $M_r > -20.5~\mathrm{mag}$ and relative velocities within $5000\,\mathrm{km\,s^{-1}}$. The density is defined as $\Sigma_5 = 5/(4\pi r_p^2)$, where $r_p$ is the projected distance to the 5th neighbor.
Additionally, to probe the influence of massive companions such as brightest cluster galaxies (BCGs), we compute the density using the 1st-nearest neighbor ($\Sigma_1$). We apply a more stringent luminosity threshold of $M_r > -22.5$ with the same velocity cut, calculating the density as $\Sigma_1 = 1/(4\pi r_p^2)$.
As shown in Figure~\ref{env} and supported by the K--S tests, the ``Blue'' group shows a tentative preference for slightly underdense environments in terms of $\Sigma_5$, whereas the remaining three groups exhibit statistically indistinguishable distributions. 
In contrast, no statistically significant differences are found among the four groups in $\Sigma_1$.
These results suggest that the distinct UV–optical slopes and effective radii defining our groups are not primarily driven by local galaxy density at the probed scales.

\section{Discussion}

\subsection{Comparisons between ADGs and LRDs}
In this subsection, we present a systematic comparison between local ADGs and high-$z$ LRDs/AGNs, focusing on the ``Compact\,\&V-shape'' group identified via K-means clustering. 
We compare these populations by examining the broad-line AGN fraction and the locations in emission-line diagnostic diagrams for the ``Compact\,\&V-shape'' group, as well as the black hole–host stellar mass relation of the broad-line subsample in the full ADG sample relative to the high-$z$ LRD/AGN population.
Together, these differences indicate systematically distinct host galaxy–black hole states in the two populations.

As a large fraction of LRDs exhibit broad H$\alpha$ emission line, we examine the broad-line subsample within the ``Compact\,\&V-shape'' group for comparison. The parent ``Compact\,\&V-shape'' group contains 427 galaxies, of which 50 show broad H$\alpha$ line (BL) emission. The fraction of broad H$\alpha$ emitters (12\%) is consistent with that of the full parent sample. The small fraction indicates that the presence of broad lines is not a common feature for the SED-shape selected local ADG sample. The properties of broad line ratio in the low-$z$ ADGs significantly differ from that of the high-redshift LRD population, where galaxies with ``V-shaped'' SEDs show a much higher ($\sim$60\% or larger) broad-line fraction \citep{greene2024uncover,Hviding2025}. We caution, however, that this apparent difference may be strongly influenced by selection biases from our local sample construction.

High-$z$ LRDs often have stringent X-ray upper limits and show little to no evidence of variability (e.g. \citealt{ananna2024ApJ,zhang2025variability}). However, a recent study of a gravitationally lensed LRD shows long-term variability over $\sim$130~yr \citep{zhang2025lensed}. 
These observational properties differ from those commonly seen in typical AGNs and pose significant challenges for interpreting their physical nature.
In contrast, within the BL samples of the local ``Compact\,\&V-shape'' group, one galaxy was reported by \citet{Baldassare2016} to exhibit the fainter broad H${\alpha}$ emission in follow-up spectroscopy, possibly attributable to a transient event such as a Type II supernova rather than persistent AGN activity. In addition, six additional dwarf galaxies show photometric variability in Zwicky Transient Facility (ZTF) and \emph{WISE} data \citep{ward2022variability}.
We further cross-matched all sources with X-ray point-source catalogs from \emph{Swift}/BAT \citep{Oh2018}, ROSAT \citep{Boller2016}, and \emph{eROSITA} \citep{bykov2024srg}. Beyond the six sources already reported in X-ray–selected studies \citep{Baldassare2017,Birchall2020,Latimer2021,bykov2024srg}, two more objects are matched to ROSAT detections. 
These contrasts point to the uniformity and extreme nature of high-$z$ LRDs compared to the more diverse phenomenology of local BL ADGs.

\begin{figure}[!htbp]
    \centering
    \includegraphics[width=0.9\linewidth]{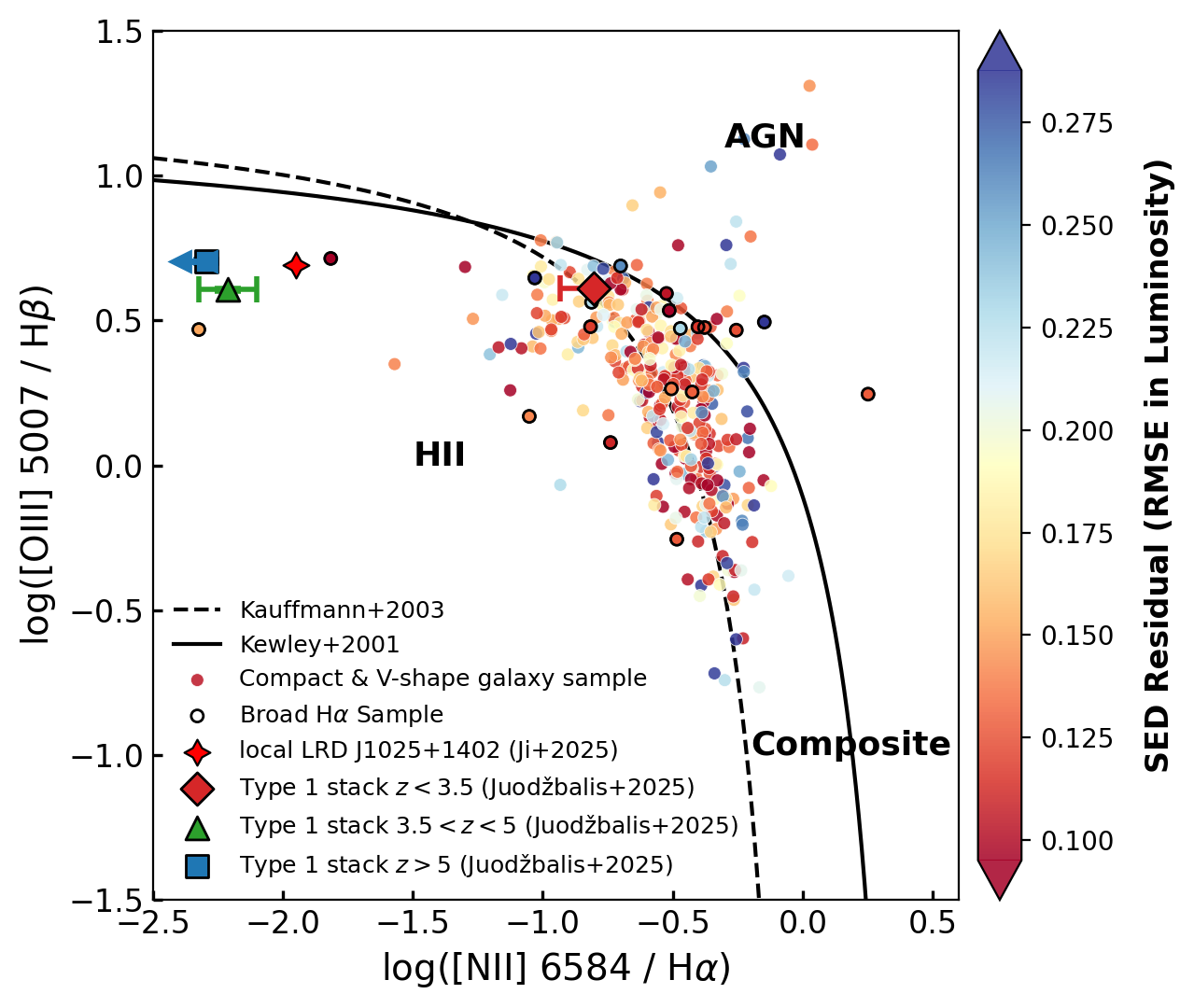}\\ 
    \caption{[N\textsc{ii}]-based BPT diagram for the ``Compact\,\&V-shape'' galaxy sample. The dashed and solid lines separate the star-forming (H\,\textsc{ii}), composite, and AGN regions \citep{Kauffmann2003,Kewley2001}. Galaxies in the ``Compact \& V-shape'' group are shown as colored dots, with the color indicating the normalized SED residual. Lower values and redder colors indicate more similar with the SED template of LRDs; the broad-line subsample is highlighted with black circles. The local LRD J1025+1402 \citep{ji2025local,lin2025local} and stacked Type 1 AGNs across $1.5 < z < 7$ from JADES \citep{Juodbalis2025} are also shown.}
    \label{BPT-diagram}
\end{figure}
\begin{figure*}[!htbp]
    \centering
    \includegraphics[width=0.8\linewidth]{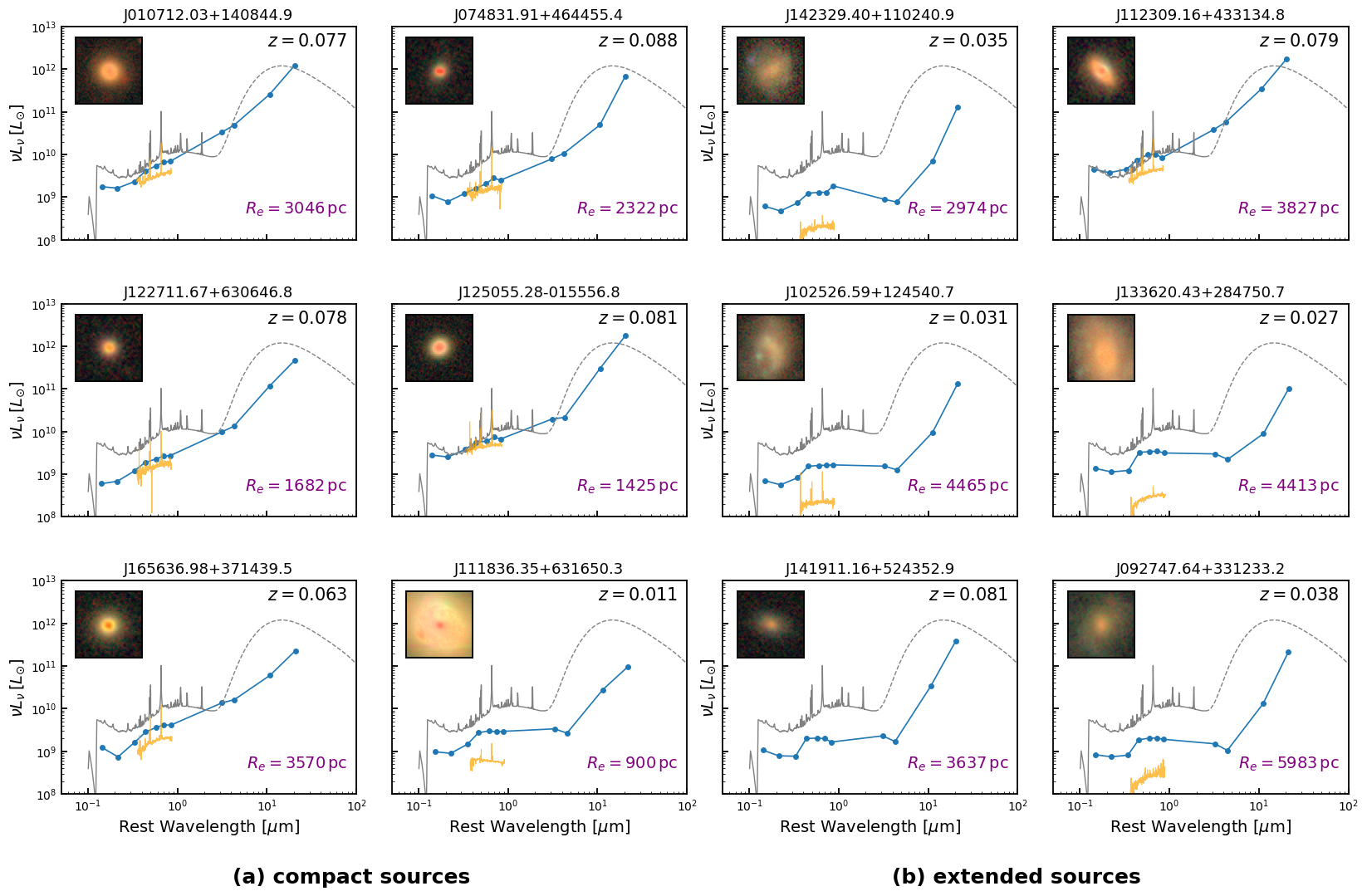}\\ 
    \caption{Example SEDs for the ``Compact\,\&V-shape'' (panel a) and ``Diffuse\,\&V-shape'' (panel b) groups. SDSS spectra are plotted in yellow, and the inset panels show DESI imaging of each galaxy. Effective radii are derived from \texttt{GALFIT} modeling, as described in Section~3.1.
    }
    \label{SED-example}
\end{figure*}

Figure~\ref{BPT-diagram} compares the narrow-line excitation properties sources in the ``Compact\,\&V-shape'' group with other broad-line AGNs (BLAGNs) using the BPT diagnostic diagram. The BL
subsample in the ADG group is marked with red circles. 
The color of each point represents the normalized SED residual in luminosity (Section~3.2), where smaller values indicate closer matches to the LRD SED template of \citet{akins2024cosmos}. 
We find that the majority of galaxies in the ``Compact\,\&V-shape'' group occupy the composite region, overlapping with the location of stacked type 1 AGNs at the low-redshift end ($z<3.5$; \citealt{Juodbalis2025}), suggesting broadly similar ionization conditions. 
Notably, two BL ADGs (J1321+3004 and J0749+4645) lie at positions comparable to the local LRD analog J1025+1402 \citep{ji2025local,lin2025local} and the stacked BLAGNs at $3.5<z<7$ from JADES \citep{Juodbalis2025}, characterized by elevated [O\,\textsc{iii}] excitation and relatively low [N\,\textsc{ii}]/H$\alpha$ ratios. 
Of these two BL ADGs, J0749+4645, which shows a smaller SED residual and a higher [N\,\textsc{ii}]/H$\alpha$ ratio, was identified via its broad H$\alpha$ emission (\citealt{Greene2007,Dong2012}; galaxy J074831.91+464455.4 in the left panel of Figure~\ref{SED-example}), whereas J1321+3004 was selected based on variability \citep{ward2022variability}.
Overall, within the local ``Compact\,\&V-shape'' group, only a small fraction of sources exhibit harder ionizing spectra and limited nitrogen enrichment, consistent with the properties observed in early galaxies, whereas the majority resemble typical low-redshift AGNs.


\begin{figure}[!htbp]
    \centering
    \includegraphics[width=0.9\linewidth]{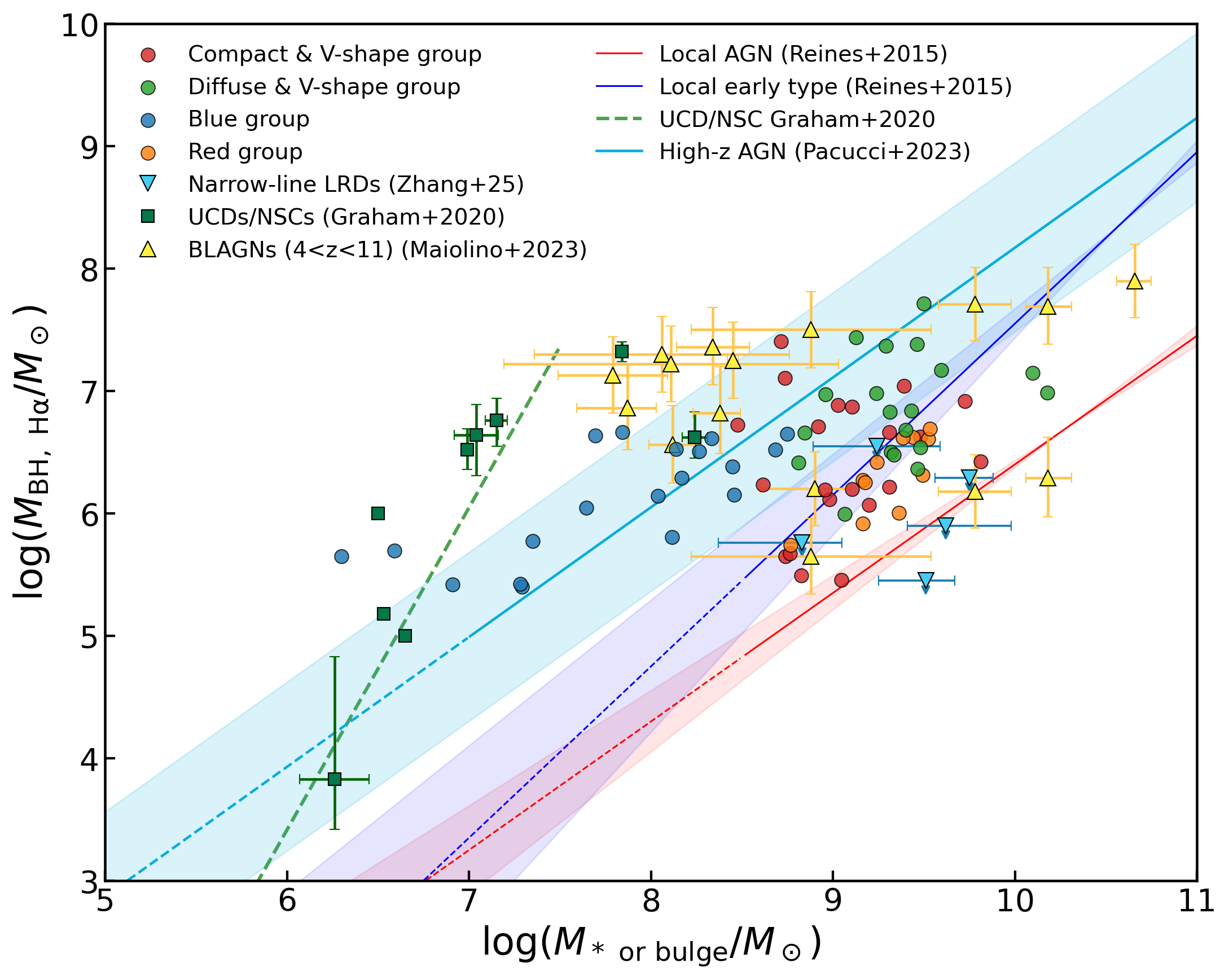}\\ 
    \caption{Black hole mass ($M_{\rm BH}$) to stellar mass ($M_\star$) relation for the four ADG groups (colored circles), high-$z$ BLAGNs (yellow triangles; \citealt{Maiolino2024}), NL LRDs (blue triangles; \citealt{zhang2025}), and local UCDs/NSCs (green squares; \citealt{Graham2020}). Various $M_{\rm BH}$--$M_\star$ scaling relations are also shown: local AGNs (\citealt{Reines2015}, red/blue lines), UCD/NSC relation (\citealt{Graham2020}, dashed green line), and high-$z$ AGNs (\citealt{Pacucci2023}, cyan line).}
    \label{Mbh-Mstar}
\end{figure}

We further compare the BL subsample of the total ADG population with \emph{JWST}-selected AGNs. Figure~\ref{Mbh-Mstar} shows the black hole mass–stellar mass ($M_{\rm BH}$–$M_\star$) relation for 70 BL ADGs with SDSS spectra, color-coded by galaxy group. We analyzed these spectra using \texttt{PyQSOFit} \citep{Guo2018}. For the ADG sample, black hole masses are estimated via single-epoch virial methods using the broad H$\alpha$ emission \citep{Reines2013}, while stellar masses are derived from \texttt{CIGALE} SED fitting (Wang et al. to be submitted). 
The high-$z$ AGN comparison sample includes BLAGNs from \citet{Maiolino2024} and the five narrow-line (NL) LRDs at $z\sim5$ reported by \citet{zhang2025}. In addition, to illustrate low-mass systems in the local Universe, we also include similar compact stellar systems such as local ultra-compact dwarfs (UCDs) and nuclear star clusters (NSCs) from \citet{Graham2020}.
As shown in Figure~\ref{Mbh-Mstar}, the BL ADG sample exhibits a large scatter, lying above or along both the local \citep{Reines2015} and high-$z$ \citep{Pacucci2023} scaling relations. These galaxies span the region between the UCDs/NSCs and the LRDs in the $M_{\rm BH}$–$M_\star$ diagram. Notably, galaxies in the ``Blue'' group occupy the low-mass end relative to the other groups, whereas galaxies with ``V-shaped'' SEDs exhibit a larger scatter toward the high-mass end. 
These different locations indicate differences in the black hole–host galaxy state both among ADG groups and between the local and high-$z$ populations across cosmic time.
We note that these offsets should be interpreted with caution, as they may partly reflect systematic differences in measurement techniques and the uncertainties in black hole and stellar mass estimates for different populations.

\subsection{Extended ``V-shape'' dwarf galaxy sample}
In addition to the ``Compact\,\&V-shape'' group, we identify another population of 267 galaxies (nearly $\sim1/4$ of the total) that display broadly similar SED shapes but with systematically larger effective radii. These systems exhibit more extended morphologies with weak or absent bulge-like components. However, the available spectra are limited by the $\sim3''$ SDSS fiber aperture, which does not capture the full spatial extent of their emission regions.
As a result, the spectra underestimate the total luminosity of these extended systems, leading to larger discrepancies between the photometric and spectroscopic luminosities in the SEDs, as shown in Figure.~\ref{SED-example} panel (b).
This subsample raises the possibility that a corresponding population may also exist at high redshift, but remains largely undetected due to their more diffuse light profiles and low surface brightness.

The bimodality in the sizes of dwarf galaxies has been reported in the literature \citep{Misgeld2011}. 
In both observational and simulation studies, compact dwarf galaxies are often interpreted as the results of mergers or tidal stripping events \citep{Watts2016,Wang2023Natur}, whereas some simulations indicate that compact dwarfs preferentially reside in lower-density environments, experience fewer major mergers, and are predicted to exhibit rounder morphologies and bluer cores \citep{DeAlmeida2024}. 
On the other hand, the diffuse dwarf galaxies with large velocity dispersions are generally attributed to substantial dark matter content, external interactions, or internal dynamical effects such as binary-star orbital motions \citep{Dotti2026}. 
Together, these studies suggest that large-scale environment may play an important role in shaping the observed size bimodality of dwarf galaxies. However, in our sample, the ``Diffuse\,\&\,V-shape'' and ``Compact\,\&\,V-shape'' groups show no statistically significant difference in environmental density or other galaxy properties, providing limited evidence for distinguishing their formation pathways. 
By contrast, only the ``Blue'' group, characterized by compact sizes, bluer colors, and lower environmental densities, appears broadly consistent with the simulation predictions.


\subsection{Clues of AGN evolution with their host galaxies}
\begin{figure*}[!htbp]
    \centering
    \includegraphics[width=0.75\linewidth]{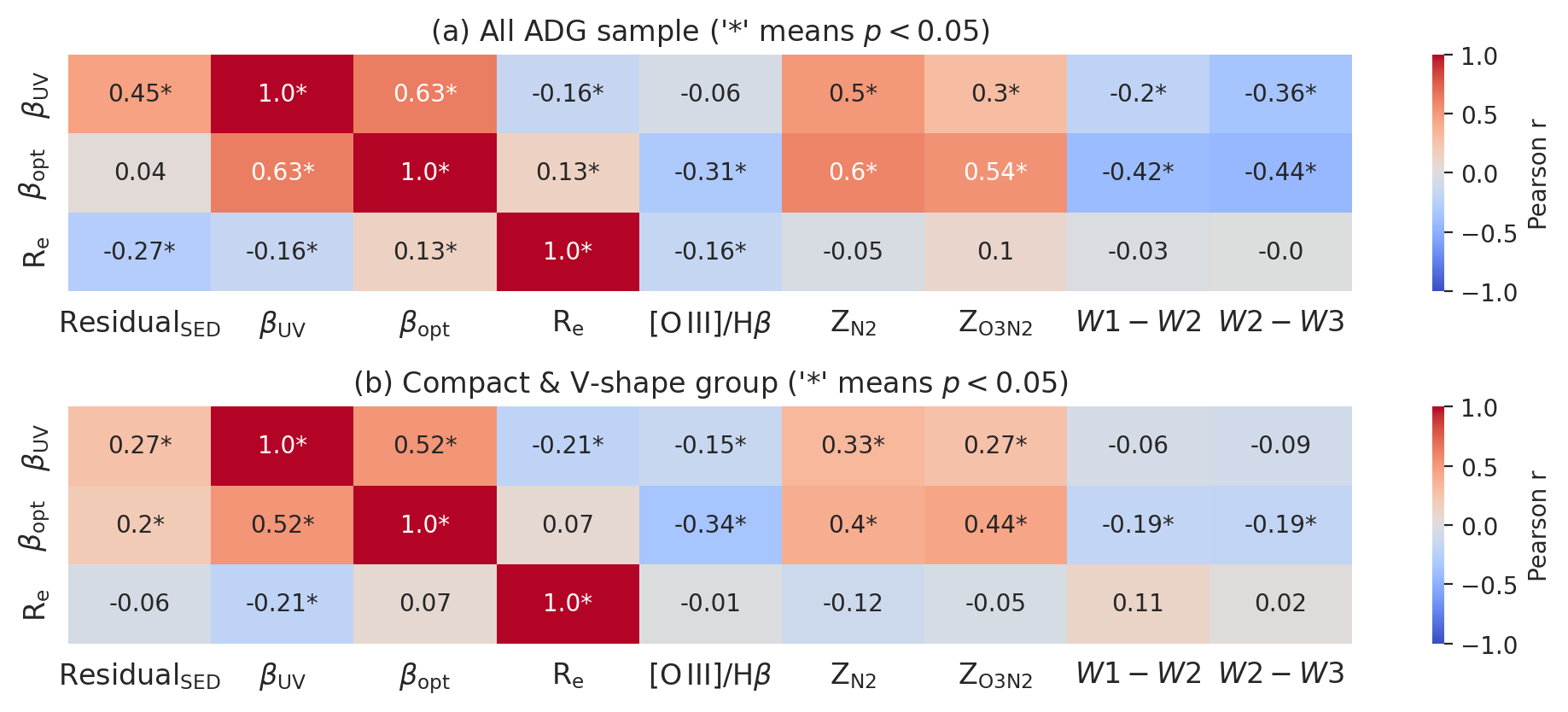}\\ 
    \caption{Correlation between the UV slope, optical slope, effective radius, and other properties of the ADG sample. (a) Pearson correlation coefficients ($r$) for various properties, including the [O\,\textsc{iii}]/H$\beta$ line ratio, gas-phase metallicity derived from the N2 and O3N2 calibrators, and WISE colors, calculated for the entire ADG sample. An asterisk (`*') indicates a statistically difference ($p < 0.05$). (b) Same as (a), but restricted to the ``Compact\,\&V-shape'' group.
    }
    \label{correlation}
\end{figure*}

Given the characteristic SED slopes and effective radii of LRDs, we investigate how these properties relate to other galaxy parameters in our local sample. We calculate the Pearson correlation coefficients between the UV slope, optical slope, effective radius, and additional properties for both the total ADG sample and the ``Compact\,\&V-shape'' group, as shown in Figure~\ref{correlation}. 
The upper panel of Figure~\ref{correlation} shows a significant anti-correlation between the UV/optical slopes and the \emph{WISE} colors. Galaxies with higher metallicities, measured using both the [N~\textsc{ii}] and [O~\textsc{iii}] calibrators, also exhibit redder slopes, consistent with the differences among the four groups discussed in Section~3.3. In contrast, the [O\,\textsc{iii}]/H$\beta$ line ratio is only negatively correlated with the optical slope and shows a weak negative correlation with the effective radius. 
Interestingly, as shown in the bottom panel, while these correlations are significant across the full sample, they become much weaker within the ``Compact\,\&V-shape'' group, except for the [O\,\textsc{iii}]/H$\beta$ line ratio. This reflects that the compact and ``V-shaped'' subsample exhibits more intense [O\,\textsc{iii}] emission relative to the entire sample.

Together with the systematic differences in galaxy properties, environment and broad-line black hole mass discussed in Section~3.3 and Section~4.3, we argue that the four identified ADG populations likely trace a sequence of evolutionary stages, with the two ``V-shaped'' groups occupying an intermediate regime between the ``Blue'' and ``Red'' groups.
This differs from the high-$z$ V-shaped LRDs, which are thought to represent an early post-formation stage \citep{Inayoshi2025,naidu2025}. The two ``V-shaped'' groups in our local sample show no significant differences in environment and most other physical properties expect for the galaxy size, implying that internal processes likely drive their morphological differences.

Given the heterogeneity of compiled local AGN samples, we estimate the number density solely from the DESI-selected BL ADG subsample within $z = 0$--0.45, which provides a more uniform and homogeneous selection. Although with a small-number sample, the seven sources in our analog group yield a number density of $\sim 2 \times 10^{-9}\,\mathrm{Mpc}^{-3}$. We emphasize that this value represents a conservative lower limit. This is because the parent ADG sample from \citet{Pucha2025} is restricted to line-emitting galaxies classified as non–star-forming on the BPT diagram. As a result, this estimate excludes AGNs hosted by quiescent galaxies lacking strong emission lines, as well as AGNs whose BPT classification falls within the star-forming region.
Compared to LRDs with the number density of $\sim10^{-5}\,\mathrm{Mpc}^{-3}\,\mathrm{mag}^{-1}$, the local sample with comparable luminosities shows a much lower number density despite similar SED shapes.

\section{Conclusion}
LRDs are known to be very compact AGN systems at high redshift and are frequently associated with local compact AGN host systems. In this paper, we investigated a large sample of local AGN-hosting dwarf galaxies (ADGs) to better understand the potential connection between local systems and the ``Little Red Dots'' (LRDs) in the high-redshift Universe.

These local galaxies exhibit typical luminosities and black hole masses comparable to those of high-$z$ LRDs, \textcolor{black}{but have systematically larger effective radii}. Using K-means clustering on the parameter space defined by UV/optical slopes, effective radii, and UV-to-optical colors, we classify the galaxies into four distinct groups. Two of these groups display a characteristic ``V-shaped" SED: the ``Compact \& V-shape" group (comprising nearly half of the total sample) and the ``Diffuse \& V-shape" group. The remaining two groups (the ``Blue'' and ``Red'' groups) instead show monotonically decreasing or increasing UV-to-optical continua.

All four groups display distinct physical characteristics that likely correspond to different evolutionary stages. The two ``V-shaped" groups show intermediate values in their \emph{WISE} colors, gas-phase metallicities, and star formation rates (SFRs). In contrast, the ``Blue'' group displays the reddest \emph{WISE} colors, the lowest metallicities, and the highest SFRs, while the ``Red'' group shows the opposite trends. Furthermore, a $k$-nearest neighbor analysis reveals no significant differences in the local environments of the three groups except for the ``Blue'' group, which shows tentatively less dense environments. This suggests that their divergent properties are not primarily driven by environmental effects.

The comparison with high-$z$ LRDs indicates that, although local ADGs share broadly similar SED features, they are likely at a different evolutionary stage. The ADGs have systematically larger effective radii than LRDs and show no strong preference for compact systems to host broad-line AGNs. Their locations on the BPT diagram further reveal that most of these low-mass galaxies follow the excitation conditions of typical low-$z$ systems, with only a few sources exhibiting emission line ratios similar to those of high-$z$ AGNs.

In summary, by comparing local AGN-hosting dwarf galaxies with high-$z$ LRDs, our results indicate that multiple physical properties of low-mass AGNs vary systematically across cosmic time, suggesting distinct modes of AGN and stellar population evolution.

\begin{acknowledgments}
This work was supported from the NSFC grant No. 12588202. 
\end{acknowledgments}

%

\vspace{5mm}
\facilities{GALEX, SDSS, DESI, WISE, JWST}


\software{\texttt{PyQSOFit} \citep{Guo2018}, 
          \texttt{CIGALE} \citep{cigale2019}, 
          \texttt{GALFIT} \citep{Peng2002, Peng2010}, 
          \texttt{scikit-learn} \citep{scikit2011}}





\bibliography{sample631}{}
\bibliographystyle{aasjournal}



\end{document}